# Gendered impact of COVID-19 pandemic on research production: a cross-country analysis

Giovanni Abramo[1], Ciriaco Andrea D'Angelo[2] and Ida Mele[3]

[1] *giovanni.abramo@iasi.cnr.it*
Laboratory for Studies in Research Evaluation, Institute for System Analysis and Computer Science (IASI-CNR), National Research Council of Italy (Italy)

[2] *dangelo@dii.uniroma2.it*
Department of Engineering and Management, University of Rome "Tor Vergata" (Italy)

[3] *ida.mele@iasi.cnr.it*
Laboratory for Studies in Research Evaluation, Institute for System Analysis and Computer Science (IASI-CNR), National Research Council of Italy (Italy)

**Abstract**
The massive shock of the COVID-19 pandemic is already showing its negative effects on economies around the world, unprecedented in recent history. COVID-19 infections and containment measures have caused a general slowdown in research and new knowledge production. Because of the link between R&D spending and economic growth, it is to be expected then that a slowdown in research activities will slow in turn the global recovery from the pandemic. Many recent studies also claim an uneven impact on scientific production across gender. In this paper, we investigate the phenomenon across countries, analysing preprint depositions. Differently from other works, that compare the number of preprint depositions before and after the pandemic outbreak, we analyse the depositions trends across geographical areas, and contrast after-pandemic depositions with expected ones. Differently from common belief and initial evidence, in few countries female scientists increased their scientific output while males plunged.

**Introduction**

Following the COVID-19 outbreak in China and the Far East first, Italy and Europe shortly after, and finally the Americas, governments adopted a body of emergency measures to contrast the pandemic diffusion. Among others, mobility restrictions and social distancing caused simultaneous disruptions to both supply and demand in a globalized world economy. On the supply side, reduction of labour supply because of infections, business closures and slowdown of production because of lockdowns and social distancing, caused a decrease in production. On the demand side, notwithstanding social safety nets introduced by governments, layoffs, loss of income, and worsened economic prospects caused a reduction in household consumption and private investment.

A rapidly growing number of studies investigate the macroeconomic effects of COVID-19 pandemic across countries, sectors in individual countries, as well as on a global scale (Pagano et al., 2020; Ludvigson et al., 2020; Baqaee & Farhi, 2020; McKibbin & Fernando, 2020). According to the World Bank, the massive shock of the COVID-19 pandemic and lockdown measures to contain it have plunged the global economy into the worst economic depression since World War II.[1] The negative effect on economies around the world is expected to lead to a decline in per capita income in about 90% of countries in 2020 (Djankov & Panizza, 2020), not to even mention the long-term social effects.

Due to containment measures, research activities, both public and private, have undergone a general slowdown as well, especially in those disciplines where the presence at work and close interaction with colleagues are necessary. "Levels of self-perceived productivity dropped, where dry lab scientists were much more likely to continue carrying out their work from home as expected (29% of dry lab scientists, but only 10% of wet lab scientists, reported "at least 80% productivity")" (Korbel & Stegle, 2020). At many major research universities, non-essential research was halted, "in what amounts to an unprecedented stoppage of academic science in modern memory" (Redden, 2020).

It is widely accepted in economic theory that R&D spending can lead to rates of return well above those expected on standard capital investment (Deleidi et al., 2019). It is to be expected then that a slowdown in research output will slow in turn global recovery from COVID-19.

In this work, we undertake empirical analysis of the impact of COVID-19 pandemic on worldwide research production, across macro-geographical areas and distinguishing by gender. We expect that the extent of slowdown in research activities varies across countries and over time, depending on the spread of infections, the extent of social restrictions, and timing of both. Furthermore, the adoption of smart working, especially at universities and public research institutions, alongside the shutdown of schools, caused a considerable increase in the scientists' workload at home that could impact research production differently across gender. In fact, the more extensive involvement of women in family responsibilities, mainly care for children (Schiebinger & Gilmartin, 2010), might have increased or relieved because of the presence of men at home. Specularly, men might face more distractions and an intensification of domestic responsibilities when confined to the home.

The implications of findings are twofold. First, any forecasts of the impact of research on economic recovery and growth, might be misleading if based on R&D spending during the pandemic. In fact, COVID-19 pandemic did not affect so much overall R&D spending, at least by governments (Radecki & Schonfeld, 2020), while it did affect research productivity, as we will show empirically. Second, if the pandemic unevenly affects research productivity across gender, any research performance evaluation should account for that, in order not to disfavour either gender in their careers and access to resources, and institutions with uneven gender distributions of research staff.

---

[1] https://www.worldbank.org/en/news/press-release/2020/06/08/covid-19-to-plunge-global-economy-into-worst-recession-since-world-war-ii, last accessed 25 January 2021.

The first empirical studies on the effects of COVID-19 pandemic on research activities focused on the response from researchers to address health issues to minimize its impact. While findings need to be verified at a later stage of the pandemic and in the years to come, from these very first investigations we learn that the volume of publications for this topic noticeably increased (Zang et al., 2020), while their quality seems below the quality average of other articles in the same journals (Zdravkovic, Berger-Estilita, Zdravkovic, & Berger, 2020). Differently from previous publications though, there seems to be a high degree of convergence between articles shared in the social web and citation counts (Kousha & Thelwall, 2020).

Soon after, the focus of scholars extended to investigate the effect of COVID-19 also on scientists' behaviour and on research activities other than COVID-19-related research. In terms of scientists' research behaviour, it is evident that the pandemic emergency led to substantial innovation in research collaboration and scholarly communication. The sense of urgency that has pervaded the world scientific community has generated amounts of data sharing and scientific research collaboration at levels that have never been seen before. It has been reported a speed-up of open early-stage research sharing, with a surge of depositions to such preprint archives as medRxiv and bioRxiv to foster large-scale early-stage research communication (Callaway, 2020).

The question of whether pandemic is disproportionately hurting the productivity of female scholars has been posed and empirically confirmed in different research disciplines and from different data sources: among Italian astronomy and astrophysics researchers (Deleidi et al., 2020); among neuro-immunologists (Ribarovska et al., 2021); among corresponding authors in medRxiv, but not in bioRxiv (Wehner, Li, & Nead, 2020); in the physical-sciences repository arXiv and, contrary to Wehner, Li and Nead (2020), in bioRxiv as well for the life sciences (Frederickson, 2020); in 11 preprint repositories, expanding disciplinary coverage, and especially on COVID-19-related research (Vincent-Lamarre, Sugimoto, & Larivière, 2020). Also, early journal submission data suggest that COVID-19 is disproportionally plunging women's research production.[2]

Given that submission data are not publicly available, in the present study we recur to preprint repositories as data sources. In the past few years, there has been a significant uptake in posting preprint works in such repositories, to accelerate the diffusion of new knowledge. Considering this, and differently from other studies on the subject, we measure the variation in research production during the pandemic, by comparing research production in the pandemic period with the expected one, as extrapolated from the trends, rather than with that in previous period.

The paper unfolds as follows. In the next section we present methods and data. In the third section we present the results of the empirical analysis. The final section is devoted to the discussion and conclusions.

**Data and methods**

Initially, we have examined the appropriateness of a number of preprint archives to meet our objectives and methodology. We discarded all of them but arXiv or bioRxiv. The reason was that either they are too recent to be able to extrapolate the trends of posting, or the volume of preprints is too small to assure robust elaborations. arXiv is a free online archive and distribution service for unpublished preprints primarily for research in physics, astronomy and mathematics. bioRxiv is the equivalent in the life sciences. We finally chose bioRxiv as the only data source, because arXiv provides authors' affiliations of a very small share of preprints (around three percent), which makes it unsuitable for country-level analyses like the one we intend to conduct.

---

[2] https://www.insidehighered.com/news/2020/04/21/early-journal-submission-data-suggest-covid-19-tanking-womens-research-productivity, last accessed 25 January 2021

bioRxiv was launched in 2013 by Cold Spring Harbor Laboratory, a not-for-profit research and educational institution,[3] recently obtaining support by the Chan-Zuckerberg initiative (Callaway, 2017). First depositions occurred in November 2013.

Articles are not peer-reviewed before being posted online on this archive but undergo a basic screening process for checking non-scientific content and plagiarism. Generally, an article may be posted prior to, or concurrently with, submission to a journal so that the authors are able to make their findings immediately available to the scientific community and receive feedback on draft manuscripts.

Before the pandemic outbreak, very few bibliometric studies had used this platform as a data source. Tsunoda, Sun, Nishizawa, Liu and Amano (2019) investigated the evolution of a set of papers posted on bioRxiv and then published in academic journals. Fraser, Momeni, Mayr and Peters (2019) investigated the citation and altmetric advantage of depositing preprints to bioRxiv. Kenekayoro (2020) recently showed that although the platform is not yet mature enough for reliable analyses, the exponential growth in preprint depositions suggests that this data source will be soon a valuable resource for discovering interesting trends on emerging or dying research fronts.

To observe the effects of the COVID-19 pandemic and of the consequent containment measures on the production of novel scientific knowledge at a short temporal distance from its outbreak, biorXiv appears particularly appropriate. Even assuming editors' acceptance rates unchanged, the use of a traditional bibliometric platforms (such as Scopus, WoS, Google Scholar, Dimension or the like) would in fact require a longer time window considering the average publication time of an article in a journal, and its indexing in bibliographic repertories.

Moreover, bioRxiv provides free and unrestricted access to all preprints posted on the server. This applies also to machine analysis of the content. Metadata is made available via a number of dedicated RSS feeds and APIs resources. For the purpose of this research, we used a wget script for retrieving all publication metadata in XML format. Data extraction took place on 16 December 2020. After retrieving all XML files related to the original deposition (version 1)[4] of all preprints on the bioRxiv server, we implemented a parser in Python for extracting relevant information from each XML file. More in detail, we extracted: date of deposition, doi (record digital identifier in bioRxiv), author first names and last name, author position, corresponding author, institution name and country, and subject area.

Since the full set of processed XML files from bioRxiv is deposited each month with delivery completing typically in the first days of the subsequent month, we can be confident that the dataset contains all depositions up to 31 November 2020.

The full retrieved dataset is made of 106,050 preprints, showing a quadratic growth along years up to the 2020 pandemic: 77 depositions in 2013, 848 in 2014, 1704 in 2015, 4590 in 2016, 11191 in 2017, 20512 in 2018, 29018 in 2019, and 38110 in 2020. The overall distribution by bioRxiv subject area is shown in Table 1.

For the identification of the gender of each author we queried the "Gender API" platform[5] by the "first name"+"affiliation_country" pair. Since the level of standardization of bioRxiv retrieved data was not very high, some initial effort was needed for manually cleaning and reconciling fields, mainly for removing umlauts and other non-ASCII characters in first names as well as for reconciling country names.

---

[3] https://www.biorxiv.org/about-biorxiv

[4] Authors can deposit a revised version of an article at any time prior to its formal acceptance by a journal.

[5] https://gender-api.com/ last accessed 25 January 2021

*Table 1. Share of total bioRxiv depositions by subject area*

| Subject area | Share | Subject area | Share |
|---|---|---|---|
| Neuroscience | 17.8% | Immunology | 3.2% |
| Bioinformatics | 9.1% | Plant Biology | 3.2% |
| Microbiology | 9.0% | Developmental Biology | 3.0% |
| Genomics | 6.1% | Systems Biology | 2.5% |
| Evolutionary Biology | 5.9% | Bioengineering | 2.4% |
| Cell Biology | 5.2% | Animal Behavior and Cognition | 1.5% |
| Genetics | 4.9% | Physiology | 1.4% |
| Biophysics | 4.4% | Epidemiology | 1.4% |
| Ecology | 4.4% | Pharmacology and Toxicology | 1.0% |
| Cancer Biology | 3.6% | Synthetic Biology | 0.9% |
| Molecular Biology | 3.5% | Scientific Communication and Education | 0.7% |
| Biochemistry | 3.5% | Other | 1.3% |

**Results**

*Spatiotemporal analysis of the pandemic impact*

The bioRxiv monthly depositions in Figure 1 fit a quadratic trend up to the end of spring 2020,[6] with a peak in June 2020. After that, we observe an abrupt inversion of the curve, indicating a disruptive effect of the pandemic on research production.

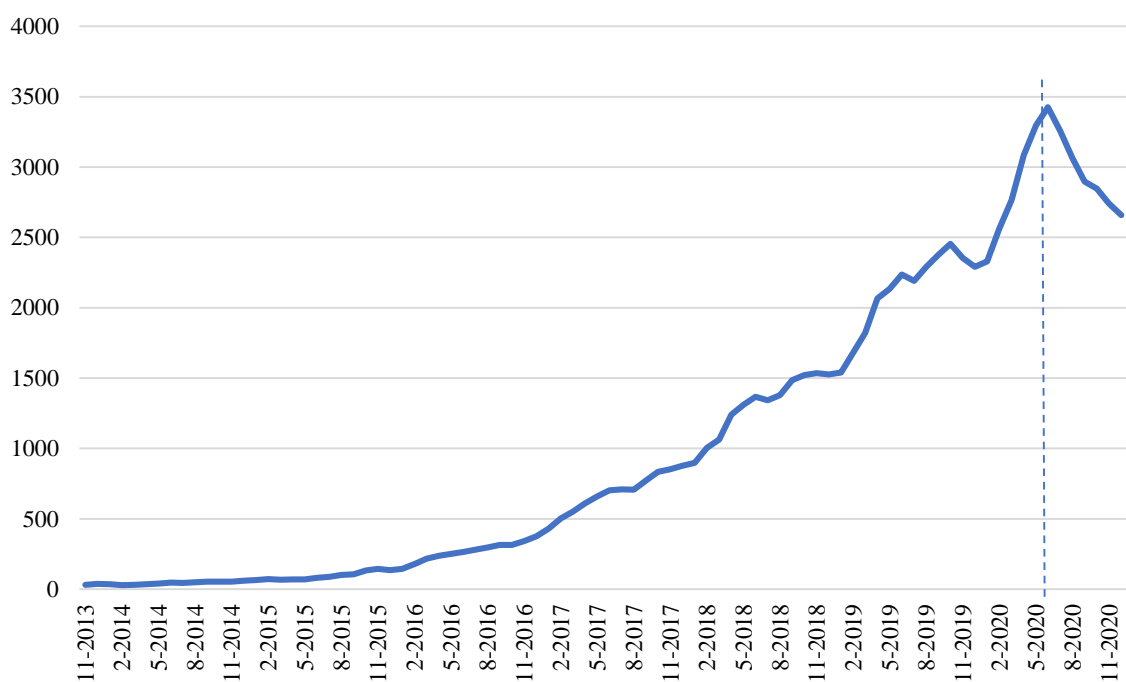

*Figure 1. Time series of overall bioRxiv preprint depositions*

To better appreciate the spatiotemporal dynamics of depositions before and after the pandemic outbreak, we stratify the data by geographical area. To start with, we identify the corresponding author of each preprint, the relevant affiliation and the corresponding geographical macro-area.

---

[6] The diagram shows the quarterly moving averages, to filter chance fluctuations.

Out of 106,050 preprints, 97,891 (92.3%) show at least one corresponding author (124,593 in total, as few publications have more than one). 101,647 corresponding authors (81.6% out of total) are provided with an affiliation[7], which can be unequivocally localized in a country and then in a geographical macro-area.

Figure 2 shows the plots of the time series depositions in three macro-areas: Europe, North America (including Canada, USA, and Greenland) and Far East (including China, Japan, Korea and Taiwan). The dot lines represent the quadratic interpolation of the yearly moving average, as measured from November 2013 to April 2020 for the Far East, and to June 2020 for Europe and North America. In order to better visualize the curve inversion in the last period, the plot starts from 2016.

The abrupt inversion of the trend is more noticeable in Europe and North America than in the Far East. The time series reflect the timing of pandemic outbreak in the different geographical areas, occurring first in the Far East but with relatively weaker effects.

It is important to notice that the average number of corresponding authors per preprint hardly changes after the pandemic outbreak (1.27 up to June 2020; 1.29 afterwards). Therefore, all trends observed with reference to the corresponding authorships remain valid for the preprints as well.

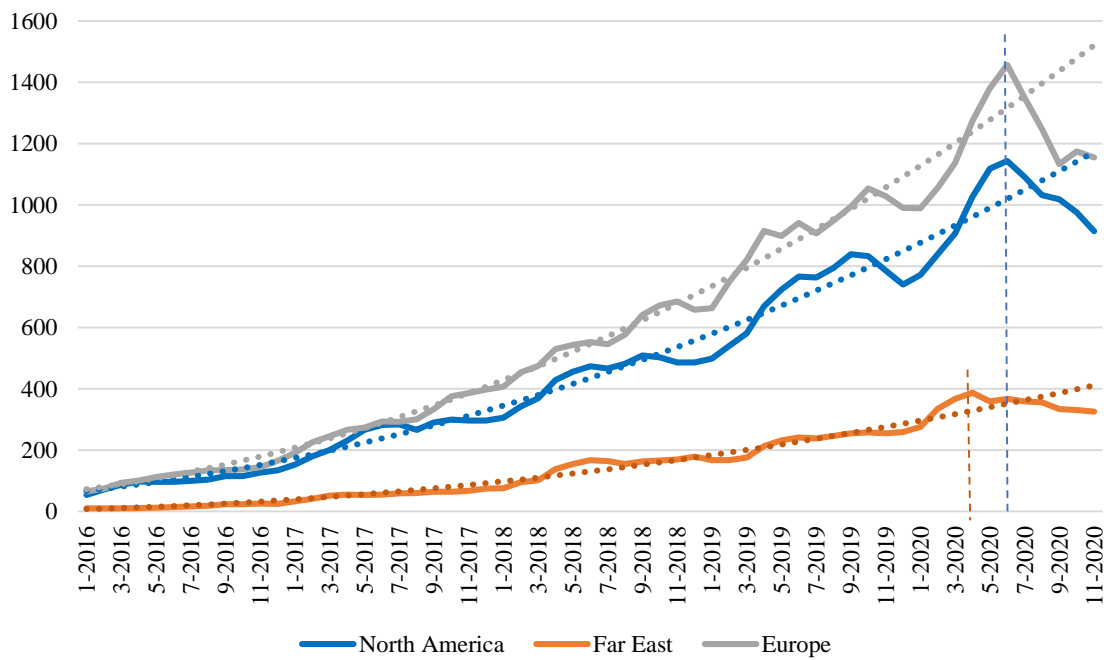

*Figure 2. Time series of bioRxiv preprint depositions by macro area of the corresponding author (2016-2020 data)*

Table 2 shows the percentage variations between observed and expected number of depositions in the period from September to November 2020.[8] The expected number of depositions is calculated as the product of the trend by the monthly seasonality coefficients derived from the time series.[9] We observe a drop in the number of depositions vìs-a-vìs the expected values of -17% at world level, with a maximum in Europe (-21%) and a minimum in the Far East (-8.8%).

---

[7] Corresponding authors with multiple affiliations are counted multiple times (155,369 total affiliations).
[8] We chose this latest subperiod, because it is less affected by the inertia of depositions related to research projects started well in advance of the pandemic outbreak.

*Table 2. Observed and expected bioRxiv preprint depositions by macro-area of affiliation of the corresponding authors (September-November 2020 data)*

|  | Observed | Expected | Variation | Confidence interval |
|---|---|---|---|---|
| North America | 2927 | 3356 | -12.8% | [-9.1%;+15.5%] |
| Far East | 994 | 1090 | -8.8% | [-10.2%;+26.7%] |
| Europe | 3523 | 4461 | -21.0% | [-5.3%;+7.4%] |
| World | 8543 | 10288 | -17.0% | [-4.0%;+7.4%] |

The recorded variation is outside the confidence interval[10] at world level, for Europe and North America, but not for the Far East.

If we consider the first author in place of the corresponding author, we obtain the plot of Figure 3, showing a pattern similar to Figure 2, the only difference being the absolute values, because in a publication there is only one first author, but there could be more than one corresponding author.

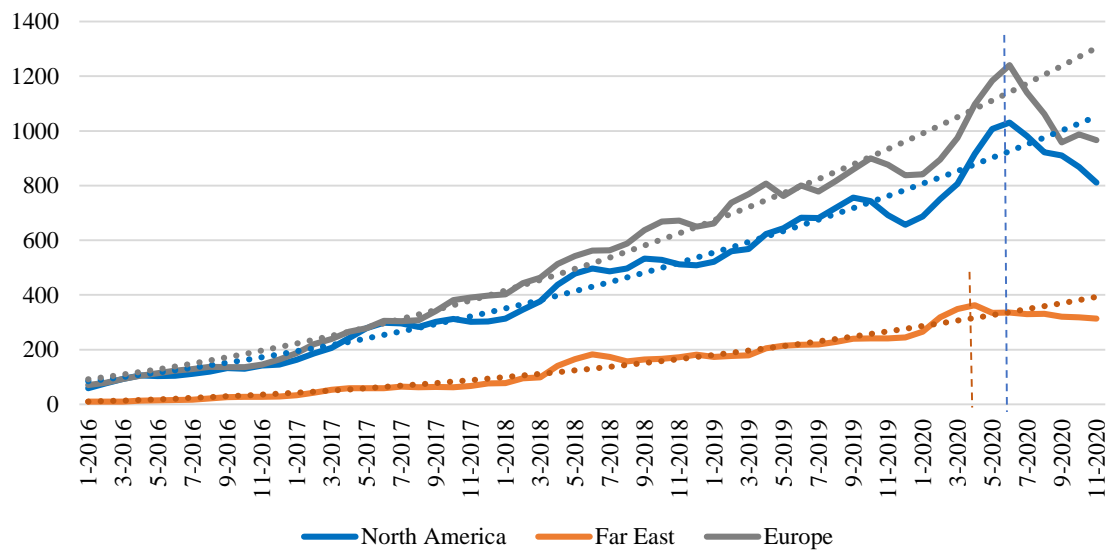

*Figure 3. Time series of bioRxiv preprint depositions by macro-area of affiliation of the first author (2016-2020 data)*

*A gender analysis of COVID-19 impact on research production*

We repeat the previous analysis further stratifying the dataset by country and gender. To discriminate gender, the first name of the author is needed alongside the country of affiliation. We discard all corresponding authors to which the Gender-API is not able to associate a gender with an accuracy equal or above 90 percent. The dataset for this analysis consists of 77,156 corresponding authors, representing 61.8 percent of 24,593 total corresponding authors in bioRxiv, and 75.9 percent of those (101,647) are provided with an affiliation that can be unequivocally localized in a country.

---

[9] With respect to the world level trend, every month presents average systematic variations, ranging from a minimum of -11.6% in December to a maximum of +5.5% in March in the overall period observed.

[10] Given by the min-max error of the estimate, i.e. the min-max variation between observed and expected values for September-November time series data.

Figures 4 to 6 show the time series for preprints posted by corresponding authors affiliated respectively to European, North American and Far Eastern research organizations. To the eye, comparing the "expected" and "observed" curves around the pandemic outbreak, it seems that the plunge in production in Europe and in the Far East is much more severe for men than for women, while it is more or less the same in North America.

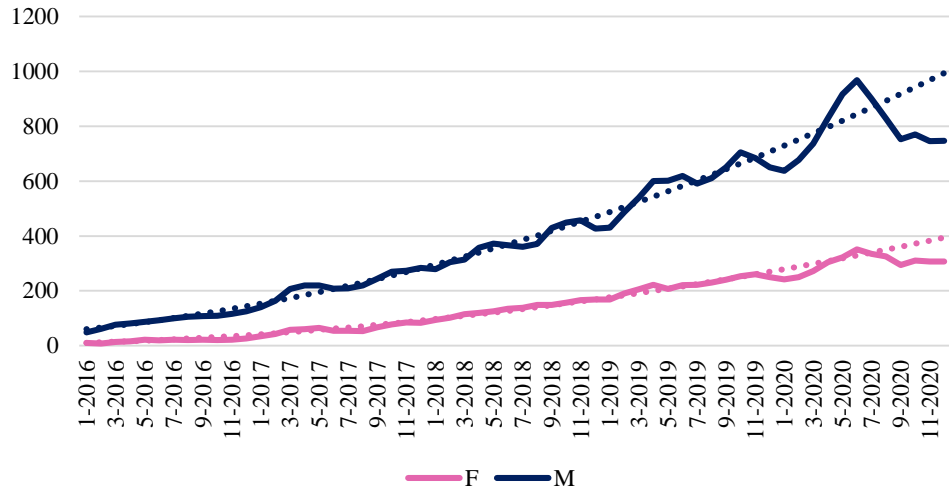

*Figure 4. Time series of bioRxiv preprint depositions by gender of corresponding authors affiliated to European research organizations*

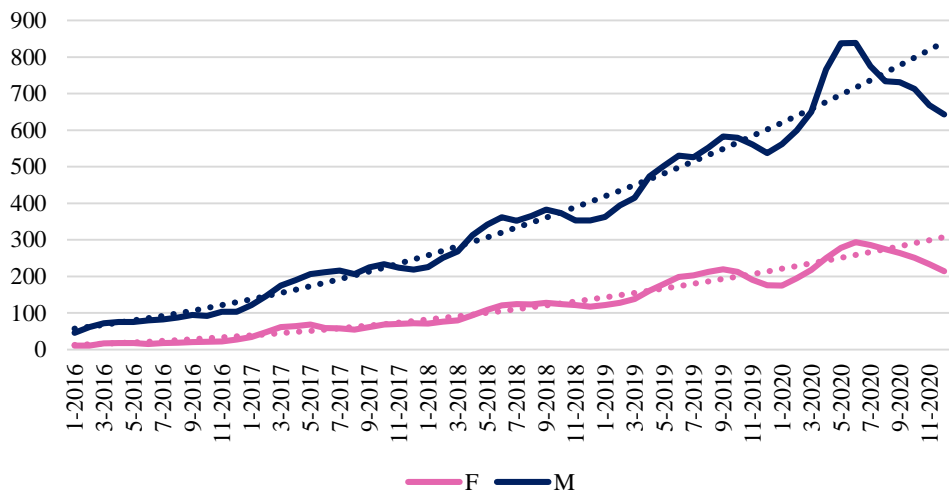

*Figure 5. Time series of bioRxiv preprint depositions by gender of corresponding authors affiliated to North American research organizations*

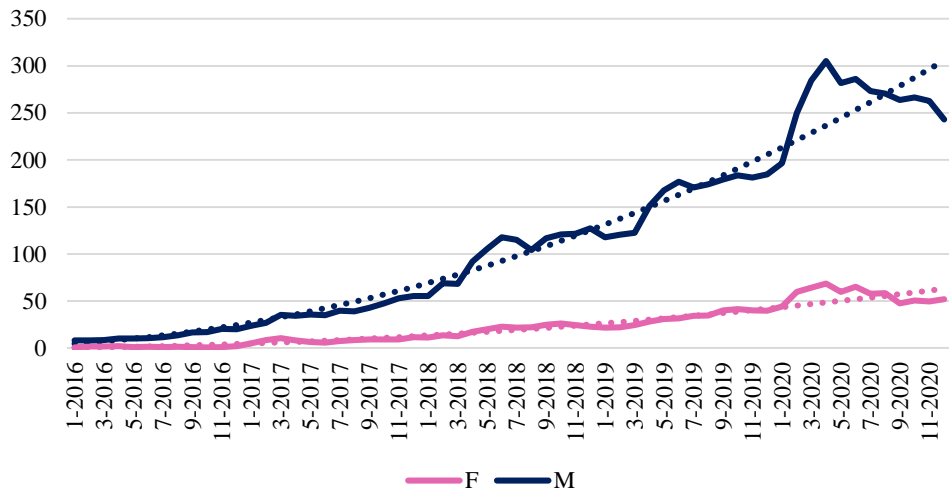

*Figure 6. Time series of bioRxiv preprint depositions by gender of corresponding authors affiliated to Far East research organizations*

A more in-depth analysis though, which takes into account also the seasonality of depositions, reveals that for North America (Table 3) the plunge in depositions is more conspicuous for women (-12.0%) than for men (-8.9%), similarly to the Far East (-6% vs +1.6%). In Europe instead, it is men who experienced a worse decrease in depositions (-18.8% vs -17.0%).

*Table 3: Observed and expected bioRxiv preprint depositions by macro-area and gender of corresponding authors (September-November 2020 data)*

|  | Gender | Observed | Expected | Variation |
|---|---|---|---|---|
| North America | F | 752 | 855 | -12.0% |
|  | M | 2139 | 2347 | -8.9% |
| Far East | F | 152 | 162 | -6.0% |
|  | M | 799 | 786 | +1.6% |
| Europe | F | 931 | 1121 | -17.0% |
|  | M | 2310 | 2845 | -18.8% |

We delve the analysis at country level considering the U.S., China, and the top eight European countries per number of depositions. Findings in Table 4 confirm the unbalance showed in Table 3 in the U.S., where female scientists reduced the output of their research activities at a higher rate than their male colleagues, -13.6% vs -9.3%. The same occurs in China, but to a less extent: -5.5% for females vs -3.0% for males. In Europe, contrasting evidences emerge. In France, Italy, Netherlands, and Switzerland it is women who are hurt more, while in Germany and Spain the opposite holds true. Quite surprisingly, in both countries female scientists raised their depositions with respect to the expected ones, respectively by 10 percent and 45 percent, while males decreased theirs by 15 percent and 9 percent. In Sweden and U.K., gender differences are hardly noticeable. It must be said that the more fine-grained the analysis the less robust the results, because of the lower number of observations. Nevertheless, what emerges at continental level is often untrue at country level, where the interplay of different containment measures, women's role in society, and family-related infrastructure unveil quite different realities.

*Table 4: Observed and expected bioRxiv preprint depositions by gender of the corresponding authors in the U.S., China, and the top eight European countries per number of depositions (September-November 2020 data)*

| Country | Gender | Observed | Expected | Variation |
|---|---|---|---|---|
| United States | F | 665 | 769 | -13.6% |
|  | M | 1919 | 2117 | -9.3% |
| China | F | 111 | 117 | -5.5% |
|  | M | 402 | 414 | -3.0% |
| France | F | 123 | 168 | -26.6% |
|  | M | 276 | 344 | -19.8% |
| Germany | F | 199 | 181 | +9.9% |
|  | M | 509 | 596 | -14.6% |
| Italy | F | 34 | 47 | -28.4% |
|  | M | 98 | 110 | -10.8% |
| Netherlands | F | 39 | 60 | -35.5% |
|  | M | 136 | 147 | -7.3% |
| Spain | F | 73 | 50 | +45.4% |
|  | M | 127 | 140 | -9.1% |
| Sweden | F | 32 | 45 | -28.5% |
|  | M | 64 | 92 | -30.6% |
| Switzerland | F | 47 | 65 | -27.7% |
|  | M | 184 | 188 | -2.1% |
| United Kingdom | F | 214 | 309 | -30.7% |
|  | M | 507 | 734 | -31.0% |

**Conclusions**

The human impacts of COVID-19 infections, and pandemic-related limitations and impediments are vast. These include disruptions to researchers that, as we showed, differ by nation and gender. In this work, we have investigated the impact of pandemic on research output. Results confirm what was expected that is a general plunge in preprint depositions, following the pandemic outbreak, consistent in timing across geographical areas: in China and the Far East first, in Europe and North America then. Probably, we will never see the consequences of the slowdown in scientific production caused by COVID-19 pandemic, in terms of published articles, as journal editors can simply raise their acceptance rates to keep the volumes unchanged. Most likely, we will witness lower average quality of publications. Evidence of that can be assessed in the years to come. Editors though might give now a precious contribution to scholars in the field by providing them with data on submission variations after the pandemic outbreak.

Contrary to what most people might expect, and early studies have announced, that female scientists are hurt more by pandemic due to the increase in family care workload, we could observe that this holds true at world level while significant exceptions occur at country level. The important lesson to be learnt is that world level analyses often hide significant differences across countries, especially when country-specific variables play a significant role in determining the outcomes.

Why gender differences occur across countries would require further investigation by scholars knowledgeable about the single country under observation. As for Italy, results are not surprising to us. In Italy women's relative share of involvement in family responsibilities, mainly care for children but also for parents and parents-in-law, is more extensive than in average EU countries. In Italy, for example, only 24 percent of the children go to kindergarten,[11] not allowing parents to be full-time occupied (Istat, 2019). Percent of population ages 65 and older (often associated with high levels of morbidity) is 24 in Italy, among the highest in the world (PRB, 2020). According to a recent survey, the majority of Italians totally agreed on the statement: "The most important role of a woman is to take care of her home and family" (EU, 2017).

---
[11] Pandemic containment measures spared kindergarten, which remained open most of the pandemic period.

Findings might be of interest to scholars in scientometrics, in the economics of innovation, and in sociology. The pandemic effects on research can inform policy makers when dealing with economic forecasts, gender equality issues, research evaluation exercises, and the assessment of the effectiveness of relevant policies and initiatives.

We appreciate several limitations embedded in our study. The field of analysis is limited to the life sciences, therefore findings and conclusions cannot be generalized to other disciplines. Data extraction was conducted during the pandemic, whose expiration is hopefully expected to occur in the next few months. Therefore, the extent of the effects that we tried to grasp is to be confirmed by future updates. It is the intention of the authors to extend the period of investigation to June 2021, in order to provide the ISSI conference participants with up-to-date and more robust findings.

Future research might entail investigation on the pandemic impact on research collaboration behaviour. It is to be expected in fact that intra-muros collaborations must have lost their advantage over extra-muros, as physical presence and personal contacts were inhibited by containment measures.